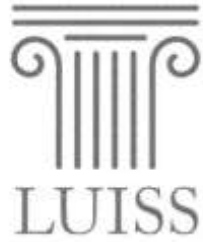

LUISS Guido Carli

# European banking supervision, the role of stress test.
*Some brief considerations.*

December 2016

Market Law and Regulation

Simone Manduchi
www.manduchi.net

# Table of Contents

# Historical perspective; a changing environment.

In the last century, Globalization heavily conditioned the world's transformation. Drastic changes affected also the economic sphere: until the XV century, Feudalism, with fragmented economic agents and economic areas, was the form. Then, it was overtaken by Mercantilism and only in the middle of XVIII century the economic system and financial environment evolved in Industrial Capitalism; that further evolved in Modern Capitalism, where one of the key aspects is represented by the ease of movements of capital, goods and people in a unified and interconnected global market which should provide – in some senses – the prosperity of the system and the achievement of market efficiency. The interconnections and the free movements of economic related elements are implicit in the capitalistic system and sometimes are safeguarded by the legislation. For instance, the European Union aims to create a Single Market (Art. 26, Treaty on the Functioning of the EU: TFEU) protecting the "*free movement of goods, persons, services and capitals*" (TITLE IV, TFEU; Art. 49 *et seq.*). All economic aspects, in fact, are affected by globalization: from human capital, to real goods, passing through financial assets; that in the last years reached 4 times the nominal world's GDP (Illustration 1). Moreover, it can be noticed (Figures 1,2,3) how links in the financial sector increased both in the core of the network, but also in periphery. EU-zone countries, together with United States, Canada and Japan, are the core of the network and they play a key role in the functioning of the entire global economic system.

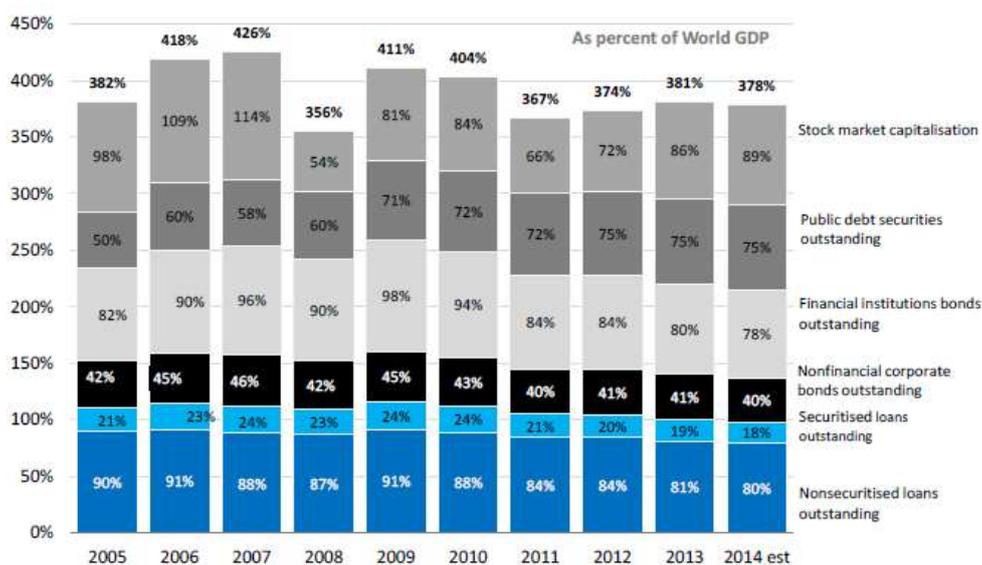

*Illustration 1: Financial assets as % of global GDP.*
Source: McKeynsey Global Institute.

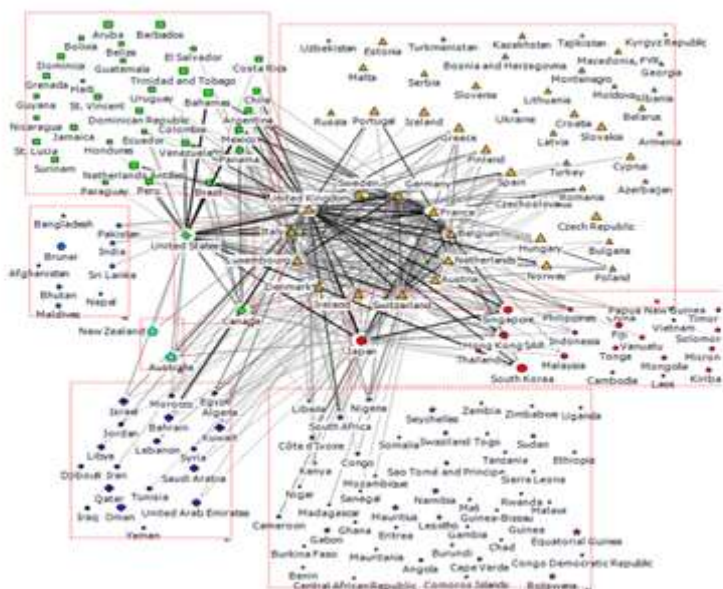

*Figure 1: Network of the cross-border banking system, 1980.*

*Source: Minoiu, Camelia, and Javier A. Reyes, 2011, "A Network Analysis of Global Banking: 1978–2009," IMF Working Paper 11/74 (Washington: International Monetary Fund).*

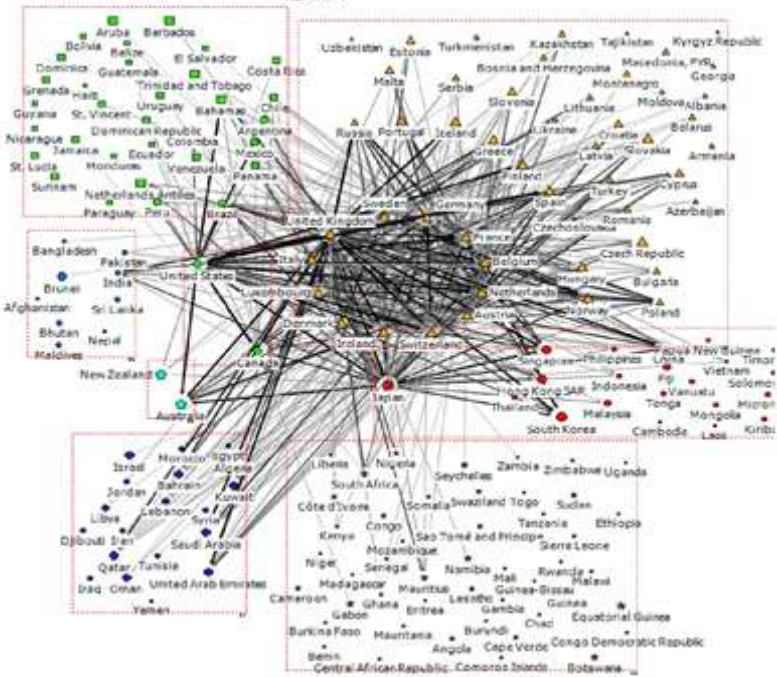

*Figure 2: Network of the cross-border banking system, 2007.*

*Source: Minoiu, Camelia, and Javier A. Reyes, 2011, "A Network Analysis of Global Banking: 1978–2009," IMF Working Paper 11/74 (Washington: International Monetary Fund).*

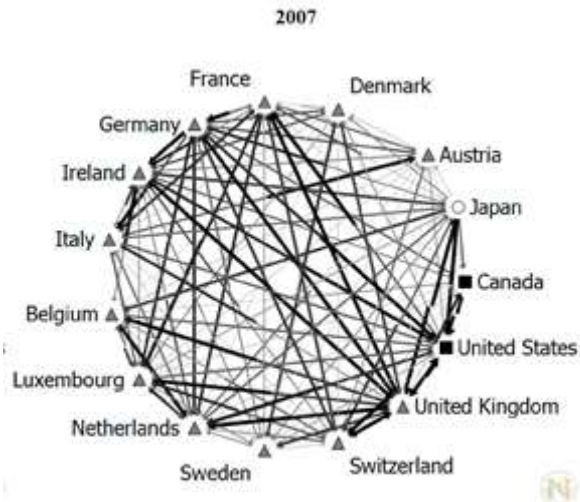

*Figure 3: Core of the network of the cross-border banking system, 2007.*

Source: Minoiu, Camelia, and Javier A. Reyes, 2011, "A Network Analysis of Global Banking: 1978–2009," IMF Working Paper 11/74 (Washington: International Monetary Fund).

# The European financial supervision

From the sub-prime crisis started in 2007 that caused a global recession, it was clear that national regulators, and supranational entities, like the European Union in the Eurozone, should safeguard the financial system and insure its stability. Investors should be able to rely on the system; as from their trust depends the entire efficiency of the market. After the crisis of 2007-2008, the De Larosière report of 2009, addressed to the European Commission, tried to delineate the framework of reforms that were needed to update, in the light of the recent events, the previous system designed by the Lamfalussy report, originally developed in 2001. The De Larosière report, suggested to substitute the third level of independent advisory groups, namely the Committee of European Banking Supervisors (CEBS), the Committee of European Security Regulators (CESR) and the Committee of European Insurance and Occupational Pensions Supervisors (CEIOP), with the so-called European Supervisory Authorities (ESAs): European Banking Authority (EBA), European Securities and Market Authority (ESMA) and the European Insurance and Occupational Pensions Authority (EIOPA). Finally, crucially for the aspects analyzed in this paper, there was the creation of the European Systemic Risk Board (ESRB), whose role is to assess and contribute to macro-prudential analysis and help to prevent systemic risk and contagion effects; which are very likely in periods of turmoil, especially for the interconnectedness shown before (Figure 3). The ESRB, together with the ESAs, the Joint Committee of the ESAs, and the national authorities, constitute the European System of Financial Supervision (ESFS). The ESFS shall put together the micro-prudential supervision made by the ESAs and the macro-prudential conducted by the ESBR.

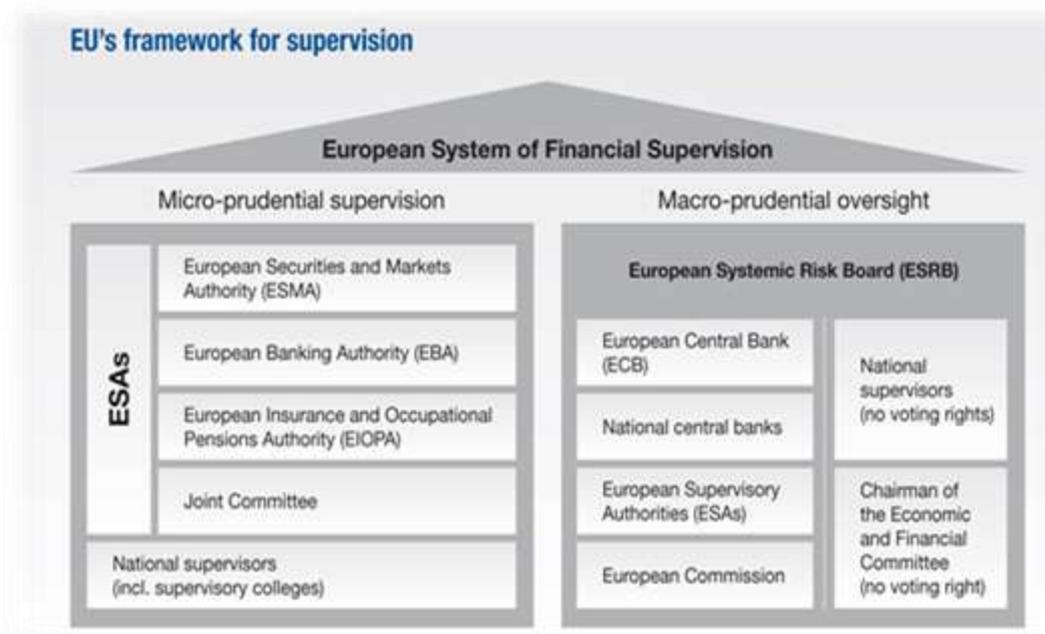

*Illustration 1: European System of Financial Supervision.*

Source: Finanssivalvonta, authority for supervision of Finland's financial and insurance sectors. http://www.finanssivalvonta.fi.

## Further developments in EU supervision; the Banking Union

ESFS coordination was not enough to avoid the crystallization of markets. For this reason, the European Commission, in 2012, proposed the Banking Union, whose aim is to create a safer environment, founded in the single rulebook *i.e.* a collection of prudential law that include mainly the Capital Requirements Regulation (CRR), the Capital Requirements Directive (CRD IV), the Bank Recovery and Resolution Directive (BRRD) and the Deposit Guarantee Schemes Directive (DGSD).

Trying to realize the Banking Union, it was introduced with Reg. 1024/2013 the Single Supervisory Mechanism (SSM), with Reg. 806/2014 the Single Resolution Mechanism (SRM) and the Deposit Guarantee Scheme (DGS), reviewed in 2014 from the original Dir. 94/19/EC.

The SSM is composed by the European Central Bank (ECB) and national authorities; it has power to "*[i] conduct supervisory reviews, on-site inspections and investigations [ii] grant or withdraw banking licences [iii] assess banks' acquisition and disposal of qualifying holdings [iv] ensure compliance with EU prudential rules [v] set higher capital requirements ("buffers") in order to counter any financial risks*". It supervises 129 banks that hold more than the 80% of banking assets in Euro area (ECB, 2016).

The logical follow-up, the SRM, aims to ensure that issues regarding distressed and failing banks are solved in the quickest way and with the lowest economic cost possible. Finally, the DGS was created to ensure that – in case of distress – up to 100 000€ are paid to depositors.

# The Basel accords
## From Basel I to Basel II

Basel agreements[1] are voluntary accords on banking supervision proposed by the Basel Committee on Banking Supervision (BCBS). Each country or entitled institution has then to implement them; the first agreement was Basel I of 1988 and it was created to give common methodologies and common rules for capital adequacy. Notably, with Basel I was given a common rule for credit risk and prudential regulation. The regulatory capital was composed by TIER 1 like stock issues, non-distributed earnings and declared reserves, and TIER 2 as all other capitals like hybrid instruments, investment assets, long term debt with maturity greater than 5 years etc. The proposed assessment of credit risk was based on the Risk Weighted Asset (RWA); in a similar way to a weighted sum of the bank's assets, different financial instruments carry different risk weights. They were organized in 5 categories: 0% (*e.g.* cash or treasuries), 20% (*e.g.* MBS with AAA rating), 50% (*e.g.* some municipalities bonds) or 100% (*e.g.* corporate bonds); in addition, there were assets with no rating. Once computed the RWA, banks with international activities had to hold, as capital, at least the 8% of the RWA[2]. Dividing the RWA by the Tier 1 Capital, is possible to obtain the Common Equity Tier 1 (CET1), which still is often used as a capitalization measure. Under BCBS lines – starting from the amendment of 1996[3] – not only Credit Risk should have been considered, but also Operational and Market Risk. To capture the latter, the concept of Value at Risk (VaR) was used. The VaR at x% is, simply, the x% quantile of the distribution of returns; it is "*the worst loss over a target horizon that will not be exceeded with a given level of confidence*"[4] (Jorion, 2007). Notably, in the legislation the probability was set at 1% and the horizon at 10 days; with at least one year of data available. However, VaR is not a good measure of risk since it does not respect subadditivity[5], even if it helped in starting to give a glance about what – in the future – would have been stress tests; in fact, the question to which VaR answers is "what happens if we are in the worst case" *i.e.* the negative tail of the distribution.

---

[1] Similar milestones were applied in Europe with Solvency II in the insurance sector (Dir. 2009/138/CE) and for IORP in the pension fund (Institutions for Occupational Retirement Provision Dir. 2003/41/EC);
[2] Total capital ratio 8%, Tier 1 capital at 4%;
[3] It was also introduced a Tier 3 capital for market risk that was some years after removed;
[4] Formally, $VaR(p)$ is that value such that $p = \int_{-\infty}^{VaR(p)} f(q)\,dq$ where $f(q)$ is the probability density function of the returns;
[5] i.e. $\varphi(X+Y) \leq \varphi(X) + \varphi(Y)$;

VaR models were widely used also in Basel II, which was published in 2004 and implemented in 2008.

Basel II proposed a three pillars model with:

1) <u>Minimum capital requirement</u>; with a new coefficient of solvability including, apart from credit risk, also market risk and operational risk. Moreover, banks could decide to use a standardized or a personalized/internal model (for instance with Internal Rating-Based approach, IRB) for measure risk. With IRB the bank had to estimate the Probability of Default (PD), Loss Given Default (LGD), Exposure At Default (EAD), which is the line of credit used by the debtor in case of insolvency, and Maturity. All these variables should be included for the computation of the Expected Loss (EL);

2) <u>Supervisory review</u>; competent authorities should have significant powers in evaluating the patrimony adequacy. In addition, banks should conduct the Internal Capital Adequacy Assessment Process (ICAAP) and proper risk management functions;

3) <u>Market discipline</u>; especially regarding transparency and disclosure.

## Basel III

After the crisis in 2010-2011, Basel III was approved. It is forward looking in the sense that mandates risk assessments on specific portfolios, depending also on the macroeconomic environment. It introduces higher minimum capitals and quality capitals, distinguishing between going concern capital Tier 1 (purest: CET1), with the logic of preventing crisis, and gone concern capital with the logic of ensuring depositors and senior creditors can be repaid in case of insolvency: Tier 2 capital) limiting leverage ratios. It introduces liquidity risk and highlights the idea of stress tests under pillar 2 'risk management and supervision' (point 115; BIS, 2011). Capital requirements for derivatives, leverage limits and capital buffers (capital to require in addition to the minimum capital) were also introduced.

In Europe, it has been implemented with the Credit Requirements Directive IV (CRD IV, Dir. 2013/36/EU) and Credit Requirements Regulation (CRR, Reg. 575/2013) and growing importance is given to stress tests.

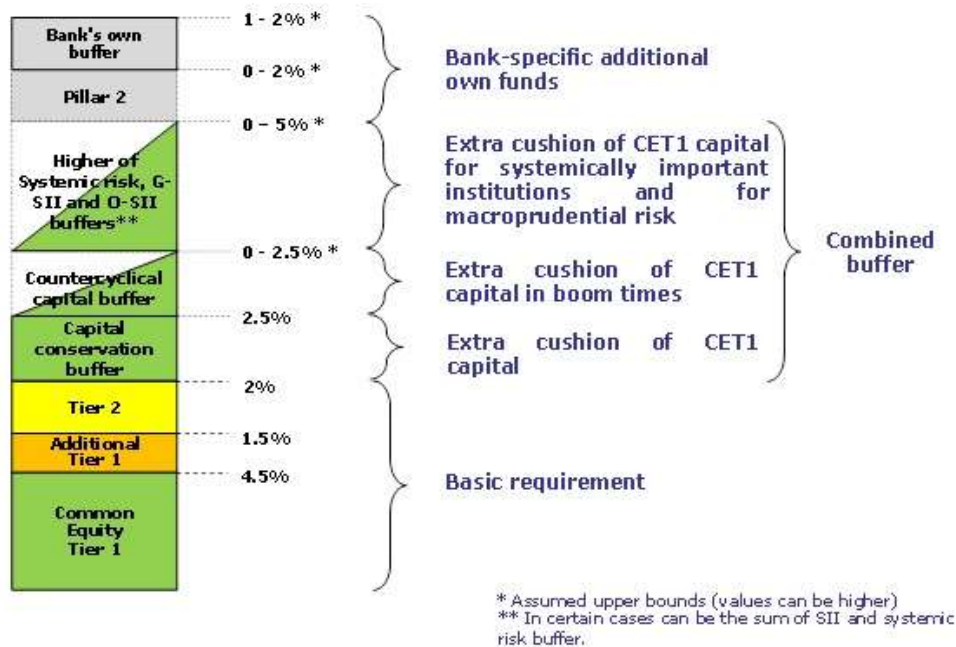

Illustration 2: Implementation of Basel III in Europe. Capital Structure.
Source: European Commission, Q&A (2013)
http://europa.eu/rapid/press-release_MEMO-13-690_en.htm

In addition to ICAAP – proposed as seen with Basel II – Internal Liquidity Adequacy Assessment Process (ILAAP) was introduced; it synthetizes the liquidity risk of a bank. ICAAP and ILAAP together with Business Model and Governance & Risk Management constitute the Supervisory Review and Evaluation Process (SREP; EBA/GL/2014/13). The SREP is a synthesis of the results of a year of analysis (Banking Supervision EU, 2016), useful for supervisors to assess how solid a bank is; even using future adverse scenarios: stress tests (point 5.6.3 of EBA/GL/2014/13). According to the results of the SREP – which should be conducted at least once a year – the competent authority can ask for supplementary capitals, a reinforcement of governance, disinvestments and all operations required to ensure the stability of the bank and henceforth the stability of the banking system.

## Stress tests

A stress test can be defined as a quantitative and statistical simulation that the supervisor or the entity conduct to assess in which way a micro agent is likely to react, in the specific scenario that has been assumed. It is a forward-looking assessment in the sense that investigates how capital and liquidity could vary in specific situations that trigger different variables and different risks; stress tests are also useful to get early warnings, trying to anticipate and limit crisis events (Papademos, 2007). In the banking sector, EBA – together with the ESRB – is allowed to use stress tests[6] in order to assess the strength of financial institutions and of market operators to adverse market developments. Stress tests can be thought as sensitivity analysis, changing one factor, or as scenario analysis, changing more factors at the same time. They are, in general, inspired to historical events and adjusted for changes in risks; they should value dependence and cyclicality with capital requirements in accordance with the two pilasters.

## Stress tests under EU law

Stress tests had been used since the last century, but only in the very last years they started to be considered in a relevant way by the regulator. The main legal provisions can be found in CRD IV, CRR and EBA guidelines. Article 98 of CRD IV describing 'Technical criteria for the supervisory review and evaluation' clearly states that "*In addition to credit, market and operational risks, the review and evaluation performed by competent authorities pursuant to Article 97 [SREP] shall include at least (...)*" and includes business model, liquidity risk, systemic risk, results from internal model regarding Title IV of CRR i.e. "own funds requirements for market risk". Liquidity and market risks shall be incorporated in the ILAAP and ICAAP and, together with business model and risk management, are the four pillars of the Supervisory Review and Evaluation Process (SREP), *ex* Art. 97 supervisory authorities shall assess and review it properly (Art. 97). But also in Article 177 of CRR, entitled 'Stress tests used in assessment of capital adequacy', it is reported that evaluations shall be conducted also bearing in mind potential stress cases:

"*(1) An institution shall have in place sound stress testing processes for use in the assessment of its capital adequacy. Stress testing shall involve identifying possible events or future changes in economic conditions that could have unfavourable effects on an institution's credit exposures and assessment of the institution's ability to withstand such changes.*

---

[6] In the U.S. has been widely used and is reported also in the Dodd-Frank act;

*(2) An institution shall regularly perform a credit risk stress test to assess the effect of certain specific conditions on its total capital requirements for credit risk. (…) The test to be employed shall be meaningful and consider the effects of severe, but plausible, recession scenarios. (…)"*

There is also an important role for accountability with a crucial role of management *ex* Article 290:

*"When evaluating solvency under stress, the shocks of the underlying risk factors shall be sufficiently severe to capture historical extreme market environments and extreme but plausible stressed market conditions. The stress tests shall evaluate the impact of such shocks on own funds, own funds requirements and earnings (…)"*

*"The results of the stress testing under the programme shall be reported regularly, at least on a quarterly basis, to senior management. (…) Senior management shall take a lead role. (…)"*

Finally, the requirements for stress test are clearly stated also in Article 100 of CRD IV:

*"(1) The competent authorities shall carry out as appropriate but at least annually supervisory stress tests on institutions they supervise, to facilitate the review and evaluation process under Article 97.*
*(2) EBA shall issue guidelines in accordance with Article 16 of Regulation (EU) No 1093/2010 to ensure that common methodologies are used by the competent authorities when conducting annual supervisory stress tests."*

On this wave, the last wide stress test was announced by EBA on 25[th] November 2015 and then launched in early 2016 covering the 70% of the banking sector: a total of 53 banks 39 of which under the SSM. It was not a pass/fail exercise since, as explained by EBA in its Q&A, the average CET1 was more than 13% and there was not an immediate need for capitalization and there were not dramatic situations. The tests included two scenarios realized by the ESRB in cooperation with EBA and ECB: a baseline scenario and an adverse scenario. In particular, the scenario had negative shocks on interest rates, exchange rates, stock price, GDP growth rate in the world and, remarkably, for EU countries (for instance in Italy was assumed a percentage change in GDP of -0.4, -1.1, 0 for years 2016, 2017, 2018), inflation/deflation, unemployment and house prices. The results – synthetized in page 33 of the EBA results report – have shown an acceptable situation. In Italy Intesa Sanpaolo was the one with the better results: 12.80% on baseline scenario and 10.21% in adverse scenario, while Mps was one of the worst with a CET1 of 12.04% in the baseline but a negative 2.23% in the adverse scenario. The National Competent Authorities has the task to check that everything is done in the proper way and according to EBA's methodology. After, these results – useful also for market transparency – were passed to competent authorities beneficial to guide targeted interventions.

# Main flaws of current stress tests

However, there are several flaws regarding stress tests. Firstly, a main issue is how stress tests are generated: they are mainly concerned with a limited number of scenarios and the corresponding assumptions. As just seen (*cfr. supra*), the adverse scenario, in the wide stress test of 2016, was only one. Furthermore, it is unknown how likely the proposed scenarios are and with which probability they will be effectively observed. In addition, they rely on macroeconomic variables that should then be converted in the microeconomic variables that directly affect the financial institution and its business; hence, other models are necessary and in many cases the links between the variables are not even clear. An example that often is made is how to translate a negative change in Gross Domestic Product (GDP) into stock returns. One may think that there is clearly a positive correlation; however, correlations are not as strong as one may think and sometimes there could be even a negative correlation between the two factors; creating another step in the statistical implementation could lead to misleading results. Moreover, it shall be pointed out that macroeconomic movements could be just a change because of the crisis and not the cause itself. Hence, the true fundamentals might not be touched by the analysis. Besides, in models there are implicit assumptions of linearity between variables and independence between phenomenons. In reality, it is much more complex; events might not be independent and a change of one variable could create endogenous risk creating a vortex that sometimes is not considered by stress tests: the world is not linear. In addition, in some cases supervisory stress tests can be conducted by banks themselves with internal models (*e.g.* IRB for ratings), introducing the issue of moral hazard: banks might be incentivized to avoid aspects where they are weaker and could tend to manipulate models to get better results (Montesi and Papiro, 2015). Finally, the severity of scenarios shall be appropriately chosen. Stress testing on very bad scenarios could – of course – show criticalities and if results are made public could lead even to the self-fulfilling prophecy that in many cases are described in the economic literature. When conducting stress testing exercises, it should be kept in mind that two variables are crucial (i) the profitability of the credit institution and (ii) the capital requirements. Injecting other capital requirements, because severe – and in some cases unlikely – stress tests suggest to do so, could not be the right choice. The main issue is to understand if the entity is profitable; keeping injecting capital in a economic-dead firm is not a proper solution and should be avoided. Especially, considered that increasing the capital reduces the profitability and returns even more.

# What could be done

Once expired the deadline for the implementation of Basel III, that currently is 2019, Basel IV will be the next step in banking regulation. It should try to amend these flaws, even if there is a lot of debate on these rules. On 24[th] March 2016, Basel Committee stated that it would have suggested to move these tests from an internal base to a validated, standardized approach. Banks and, especially, the Securities Industry and Financial Markets Association (SIFMA) harshly criticized this decision (Risk.net, 2016), giving the idea of how difficult could be to match all the interests. EBA is currently drafting the 'Guidelines on stress testing and supervisory stress testing' (EBA/CP/2015/28), *ex* art. 100 Dir. 2013/36/EU. Some aspects could be improved following the suggestions by Montesi and Papiro (2015). In particular, it shall be needed to design multiple scenarios, even with usage of Monte Carlo Markov Chain (MCMC) simulations, that will include also microeconomic variables that could change the results of the stress test, for instance also including conduct risk and giving importance to all the risks that a bank could hold. Moreover, it is crucial to think about variables having in mind stochastic distributions: variables, instead of being fixed and deterministic, shall have proper Probability Distribution Functions (PDF) and results should be used and interpreted according to where they belong in the distribution.

# References (Harvard Reference Style)